\documentclass[singlecolumn,epjc3]{svjour3}
\smartqed  
\RequirePackage{graphicx}
\journalname{Eur. Phys. J. C}

\begin{document}

\title{The Finslerian wormhole models}

\author{Farook Rahaman\thanksref{e1,addr1}
\and Nupur Paul\thanksref{e2,addr2} \and Ayan Banerjee\thanksref{e3,addr3}
\and S.S. De\thanksref{e4,addr4} \and Saibal Ray\thanksref{e5,addr5}
\and A.A. Usmani\thanksref{e6,addr6}}

\thankstext{e1}{e-mail: rahaman@associates.iucaa.in}
\thankstext{e2}{e-mail: nnupurpaul@gmail.com}
\thankstext{e3}{e-mail: ayan\_7575@yahoo.co.in}
\thankstext{e4}{e-mail: ssddadai08@rediffmail.com}
\thankstext{e5}{e-mail: saibal@associates.iucaa.in}
\thankstext{e6}{e-mail: anisul@associates.iucaa.in}

\institute{Department of Mathematics,
Jadavpur University, Kolkata 700032, West Bengal,
India\label{addr1} \and Department of Mathematics,
Jadavpur University, Kolkata 700032, West Bengal,
India\label{addr2} \and Department of Mathematics,
Jadavpur University, Kolkata 700032, West Bengal,
India\label{addr3} \and Department of
Applied Mathematics, University of Calcutta, Kolkata 700009,
West Bengal, India\label{addr4} \and Department of Physics,
Government College of Engineering and Ceramic Technology, Kolkata
700010, West Bengal, India\label{addr5} \and Department of Physics,
Aligarh Muslim University, Aligarh 202002, Uttar
Pradesh, India\label{addr6}}

\date{Received: date / Accepted: date}

\maketitle

\begin{abstract}
We present models of wormhole under the Finslerian structure of
spacetime. This is a sequel of our previous work~\cite{FR} where
we constructed a toy model for compact stars based on the
Finslerian spacetime geometry. In the present investigation, a
wide variety of solutions are obtained that explore wormhole
geometry by considering different choices for the form function
and energy density. The solutions, like the previous
work~\cite{FR}, are revealed to be physically interesting and
viable models for the explanation of wormholes as far as the
background theory and literature are concerned.
\end{abstract}


\section{INTRODUCTION}
The traversable Lorentzian wormholes, a hypothetical narrow
`bridges' or `tunnels' connecting two regions of the same universe
or two separate universes, have become a subject of considerable
interest in the last couple of years following the pioneer work by
Morris and Thorne~\cite{MT1988}. Such wormholes, which act as a
kind of `shortcut' in spacetime, are offspring of the Einstein
field equations~\cite{MT1988,Visser1995} in the hierarchy of the
black holes and whiteholes. The most striking property of such a
wormhole is the existence of inevitable amount of exotic matter
around the throat. The existence of this static configuration
requires violation of the null energy condition
(NEC)~\cite{Hochberg1997,Ida1999,Visser2003,Fewster2005,Kuhfittig2006,Rahaman2006,Jamil2010}.
This implies that the matter supporting the wormholes is exotic.
As the violation of the energy condition is particularly a
problematic issue, Visser et al.~\cite{Visser2003} have shown that
wormhole spacetimes can be constructed with arbitrarily small
violation of the averaged null energy condition. It is noted that
most of the wormhole solutions have been devoted to study static
configurations that must satisfy some specific properties in order
to be traversable. However, one can study the wormhole
configurations such as dynamical
wormholes~\cite{Hochberg1998,Hayward1999}, wormholes with
cosmological constant $\Lambda$~\cite{Lemos2003,Rahaman2007},
rotating wormholes~\cite{Teo1998,Kuhfittig2003} etc. to obtain a
panoramic representation of different physical aspects of the
wormhole structures.

Scientists have been trying to describe the wormhole structure in
two ways: either modifying the Einstein theory or matter
distribution part. In this paper, however we shall study the
wormhole solution in the context of Finsler~\cite{Bao2000}
geometry, which is one of the alternatives of the general
relativity. This involves with the Riemann geometry as its special
case where the the four-velocity vector is treated as independent
variable. As a historical anecdote we note that Cartan~\cite{Car}
initiated the self-consistent Finsler geometry model in 1935.
Thereafter, the Einstein-Finsler equations for the Cartan
$d$-connection were introduced in 1950~\cite{hor}. As a
consequence of that various models of the Finsler geometry in
certain applications of physics were
studied~\cite{vac1,vac2,Bao2015}. Though in some of the cases, the
Finsler pseudo-Riemannian configurations were considered, however,
investigators were unable to obtain any exact solution. In the
beginning of 1996, Vacuru~\cite{vac3,vac4} constructed
relativistic models of the Finsler gravity in a self-consistent
manner. He derived Finsler gravity and locally anisotropic spinors
in the low energy limits of superstring/supergravity theories with
$N$-connection structure the velocity type coordinates being
treated as extra-dimensional ones. Vacaru and his
group~\cite{Vacaru2012,Stavrinos2013,Rajpoot2015} explained the
so-called anholonomic frame deformation method (AFDM) by using the
Finsler geometry methods, which allows to construct generic
off-diagonal exact solutions in various modified gravity theories.

In this direction, numerous class of exact solutions for the
Finsler modifications of black hole, black ellipsoid/torus/brane
and string configurations, locally anisotropic cosmological
solutions have been developed for the so-called canonical
$d$-connection and Cartan $d$-connections. Therefore, it is seen
that in recent years the Finsler geometry has drawn much attention
due to its potentiality to explain various issues that can not be
explained by the Einsteinian gravity. It has been argued that
cosmic acceleration can be explained in the context of the Finsler
geometry without invoking any dark matter~\cite{Chang2008} or dark
energy~\cite{Chang2009}. Very recently Chang et al.~\cite{Chang}
have studied the kinematics and causal structure in the Finsler
spacetime and the study reveals the superluminal phenomena of
neutrinos. Pfeifer and Wohlfarth~\cite{Pfeifer} have obtained an
action for the Finsler gravity by including the description of
matter fields which are coupled to the Finsler spacetime from the
first principles. An exact vacuum solution for the Finsler
spacetime have found by Li and Chang~\cite{Li2014}. They showed
that the Finslerian covariant derivative is conserved for the
geometrical part of the gravitational field equation.

Inspired by our previous work~\cite{FR} on compact stars in the
context of Finslerian spacetime geometry, we obtain exact wormhole
solutions in this paper. We assume some definite forms of wormhole
structures and try to find out matter distributions that
reproduces it. We thus consider specific shape functions and
impose restricted choices of redshift functions for the solutions.
We study the sensitivity of our solutions with respect to the
parameters defining the shape functions. Besides, we also consider
specific energy density and dark energy equation of state,
$p_r=\omega\rho$. The sensitivities of our results for $\omega <
-1$ have also been studied. We find interesting results.

The paper is organized as follows: in  Sect. 2 we discuss the
basic equations based on the formalism of the Finslerian geometry.
Sect. 3 provides several models of the wormhole. We have also
analyzed the models in Sect. 3. The paper ends with a short
discussion in Sect. 4.

\section{Basic equations based on the Formalism of Finsler geometry}
To search  wormhole structure one needs to introduce the metric. Let
us consider the Finsler structure is of the form~\cite{Li2014}
\begin{equation}
F^2=B(r)y^ty^t-A(r)y^ry^r-r^2\bar{F}^2(\theta,\varphi,y^\theta,y^\varphi).
\end{equation}

In this study, we consider  $\bar{F^2}$ in the following form
\begin{equation}
\bar{F^2}=y^\theta y^\theta+f(\theta,\phi)y^\phi y^\phi
,\end{equation}

Thus  \[ \bar{g}_{ij}=diag(1,f(\theta,\phi)),~~~~~~~and~~~~~ \bar{g}^{ij}=diag(1,\frac{1}{f(\theta,\phi)});~~~~~[i,j=  \theta , \phi].\]

It is easy to calculate the geodesic spray coefficients $ \left[G^\mu = \frac{1}{4} g^{\mu \nu}
\left( \frac{\partial^2 F^2}{\partial x^\lambda \partial y^\nu} y^\lambda - \frac{\partial F^2}{\partial x^\nu} \right) \right] $ from $\bar{F^2}$ as
\[ \bar{G}^\theta=-\frac{1}{4}\frac{\partial f}{\partial \theta}y^\phi y^\phi,\]

\[ \bar{G}^\phi=\frac{1}{4f}\left(2\frac{\partial f}{\partial
\theta}y^\phi y^\theta+\frac{\partial f}{\partial
\phi}y^\phi y^\phi\right).\]

These yield Ricci scalar $\left(Ric\equiv R^\mu_\mu=\frac{1}{F^2}\left(2\frac{\partial
G^\mu}{\partial x^\mu}-y^\lambda \frac{\partial^2G^\mu}{\partial
x^\lambda\partial y^\mu}+2G^\lambda\frac{\partial^2G^\mu}{\partial
y^\lambda\partial y^\mu}-\frac{\partial G^\mu}{\partial
y^\lambda}\frac{\partial G^\lambda}{\partial y^\mu}\right)\right)$    in Finsler geometry
\[ \bar{F}^2\bar{Ric}=y^\phi y^\phi \left[-\frac{1}{2}\frac{\partial^2
f}{\partial \theta^2}  +\frac{1}{2f}\frac{\partial^2 f}{\partial
\phi^2}-\frac{1}{2}\frac{\partial}{\partial
\phi}\left(\frac{1}{f} \frac{\partial f}{\partial \phi}\right) -
\frac{1}{2f^2}\left(\frac{\partial f}{\partial \phi}\right)^2
-\frac{1}{4f} \left(\frac{\partial f}{\partial
\theta}\right)^2+\frac{1}{4f}\frac{\partial f}{\partial
\phi}\frac{1}{f}\frac{\partial f}{\partial \phi}+\frac{\partial
f}{\partial \theta}\frac{1}{2f}\frac{\partial f}{\partial
\theta}-\frac{1}{4f^2}\left(\frac{\partial f}{\partial
\phi}\right)^2\right]\]

\begin{equation}+y^\theta
y^\theta\left[-\frac{1}{2}\frac{\partial}{\partial
\theta}\left(\frac{1}{f} \frac{\partial f}{\partial
\theta}\right)-\frac{1}{4f^2}\left(\frac{\partial f}{\partial
\theta}\right)^2\right]+ y^\phi
y^\theta\left[\frac{1}{f}\frac{\partial^2f}{\partial \theta
\partial \phi}-\frac{1}{f^2}\left(\frac{\partial f}{\partial
\theta}\right)\left(\frac{\partial f}{\partial
\phi}\right)-\frac{1}{2}\frac{\partial}{\partial
\theta}\left(\frac{1}{f} \frac{\partial f}{\partial
\phi}\right)-\frac{1}{2}\frac{\partial}{\partial
\phi}\left(\frac{1}{f} \frac{\partial f}{\partial
\theta}\right)\right].
\end{equation}

Note that the coefficient of $y^\phi y^\theta =0$ iff, $f$ is independent of $\phi$, i.e.
\begin{equation}
f(\theta, \phi) =f(\theta),
\end{equation}
where the coefficient of $y^\theta y^\theta $ and $y^\phi y^\phi $ are non-zero.

Now, using Eq. (4) in Eq. (3), we get
\[ \bar{F}^2\bar{Ric}=\left[-\frac{1}{2f}\frac{\partial^2
f}{\partial \theta^2}  +\frac{1}{4f^2}\left(\frac{\partial f}{\partial
\theta}\right)^2\right] (y^\theta y^\theta+f y^\phi y^\phi).\]

Thus we obtain $\bar{Ric}$ as
\begin{equation}
\bar{Ric}= -\frac{1}{2f}\frac{\partial^2
f}{\partial \theta^2}  +\frac{1}{4f^2}\left(\frac{\partial f}{\partial
\theta}\right)^2,
\end{equation}
which may be a constant or a function of $\theta$.

For constant value, say $\lambda$, one can get the Finsler
structure $\bar{F^2}$ as expressed in Eq. (2) in the following
categories:
\[\bar{F}^2 = y^\theta y^\theta  + A \sin^2(\sqrt{\lambda} \theta )y^\phi y^\phi,~(for~ \lambda > 0); \]
\[ = y^\theta y^\theta  + A \theta^2 y^\phi y^\phi,~(for~ \lambda = 0); \]
\begin{equation}
= y^\theta y^\theta  + A \sinh^2(\sqrt{-\lambda} \theta )y^\phi y^\phi,~(for~ \lambda < 0).
\end{equation}

Without any loss of generality one can take $A$ as unity.

Now, the Finsler structure given in Eq. (1), assumes the following form
\[ {F}^2 =B(r) y^ty^t - A(r) y^ry^r -r^2 y^\theta y^\theta  - r^2\sin^2 \theta y^\phi y^\phi  + r^2\sin^2 \theta y^\phi y^\phi -r^2\sin^2(\sqrt{\lambda} \theta )y^\phi y^\phi.\]

That is
\begin{equation}
{F}^2 = \alpha^2 + r^2 \chi (\theta)y^\phi y^\phi,~~~~~~~~~~~~~~~~~~~~~~~~~~~~~~~~~~~~~~~~~~~~~~~~~~~~~~~~~~~~~~~~~~~
\end{equation}
where  $\chi (\theta) = \sin^2 \theta - \sin^2(\sqrt{\lambda} \theta )$ and $\alpha$ is a Riemannian metric.

Hence
\[ {F} = \alpha \sqrt{ 1 + \frac{r^2 \chi (\theta)y^\phi y^\phi}{\alpha^2}}.\]

For the choice $b_\phi= r\sqrt{ \chi (\theta)} $, we get
\begin{equation}
F = \alpha \phi(s)~~ , ~~\phi(s) = \sqrt{1+s^2},
\end{equation}
 where
 \[  s =\frac{(b_\phi y^\phi)}{\alpha} = \frac{\beta}{\alpha}, \]
\[ b_\mu = ( 0,0,0,b_\phi)  ~,~ ~ b_\phi y^\phi = b_\mu y^\mu = \beta  ~, ~ (\beta ~is ~one~ form ).\]
This indicates that $F$ is the metric of $(\alpha,~\beta)$-Finsler space.

Isometric transformations of Finsler structure~\cite{Li2014} yields the Killing equation $K_V(F) =0$ in the Finsler space as follows
\begin{equation}
\left(\phi(s)-s\frac{\partial \phi(s)}{\partial s}\right)K_V(\alpha) + \frac{\partial \phi(s)}{\partial s }K_V\beta) =0,
\end{equation}
where
   \[K_V(\alpha) = \frac{1}{2 \alpha}\left(V_{\mu \mid \nu} +V_{\nu \mid \mu} \right)y^\mu y^\nu,\]
    \[K_V(\beta) =  \left(V^\mu \frac{\partial b_\nu}{\partial x^\mu } +b_{ \mu}\frac{\partial V^\mu}{\partial x^\nu} \right)y^\nu.\]
Here $``\mid"$ indicates the covariant derivative with respect to the Riemannian metric $\alpha$.

In the present consideration we have
   \[K_V(\alpha)+ sK_V(\beta)=0 ~~or~~  \alpha K_V(\alpha)+  \beta K_V(\beta)=0.\]

This yields
 \begin{equation}  K_V(\alpha)= 0 ~~and~~K_V(\beta)=0,
\end{equation}
 or
\begin{equation} V_{\mu \mid \nu} +V_{\nu \mid \mu} =0,
\end{equation}
 and
\begin{equation} V^\mu \frac{\partial b_\nu}{\partial x^\mu } +b_{ \mu}\frac{\partial V^\mu}{\partial x^\nu} =0.
\end{equation}

Interestingly, we note that the second Killing equation constrains
the first one (Killing equation of the Riemannian space). Hence,
it is responsible for breaking the isometric symmetry of the
Riemannian space.

Actually, the present Finsler space (for the case $\bar{F^2}$ as
quadric in $ y^\theta~ \& ~y^\phi $) can be determined from a
Riemannian manifold  $( M, g_{\mu \nu}(x))$ as we have
   \[F(x,y) =\sqrt{g_{\mu \nu}(x)y^\mu y^\nu }.\]

It is to be noted that this is a semi-definite Finsler space. As a
result, we can use covariant derivative of the Riemannian space.
The Bianchi identities coincide with those of the Riemannian space
(being the covariant conservation of Einstein tensor). The present
Finsler space reduces to the Riemannian space and consequently the
gravitational field equations can be achieved. Again, following Li
et al.~\cite{LWC}, we can find the gravitational field equations
alternatively. They have also proved the covariantly conserved
properties of the tensor $G^\mu_\nu$  in respect of covariant
derivative in Finsler spacetime with the Chern-Rund connection.

It is also to be noted that the gravitational field equation in
the Finsler space is controlled to the base manifold of the
Finsler space~\cite{Li2014}, and the fiber coordinates $y^i$ are
set to be the velocities of the cosmic components (velocities in
the energy momentum tensor). It is also shown by Li et
al.~\cite{Li2014} that the gravitational field equation could be
derived from the approximation of the work done by Pfeifer et
al.~\cite{Pfeifer}. The gravitational dynamics for the Finsler
spacetime in terms of an action integral on the unit tangent
bundle has been studied by Pfeifer et. al.~\cite{Pfeifer}. Again
the gravitational field equation in the Finsler space is
insensitive to the connection because $G_\nu^\mu$ are obtained
from the Ricci scalar which is, in fact, insensitive to the
connections and depend only on the Finsler structure.

Thus the gravitational field equation in the Finsler space could
be derived from the Einstein field equation in the Riemannian
spacetime with the metric (1) in which the metric $\bar{g}_{ij}$
is given by
\[\bar{g}_{ij} = diag~(~ 1~,~~ \sin^2 \sqrt{\lambda} \theta~).\]

That is
   \[g_{\mu\nu }= diag~(~B,~-A,~ -r^2~,~~ -r^2\sin^2 \sqrt{\lambda} \theta~ ).\]

Here the new parameter $\lambda$ plays a significant role in the
resulting field equations in Finsler space and consequently
affects the Finsler geometric consideration of the wormhole
problem.

Finsler structure (1) yields geodesic spray coefficients as
\begin{equation}
G^t=\frac{B'}{2B}y^ty^r
\end{equation}
\begin{equation}
G^r=\frac{A'}{4A}y^ry^r+\frac{B'}{4A}y^ty^t-\frac{r}{2A}\bar{F^2}
\end{equation}
\begin{equation}
G^\theta=\frac{1}{r}y^\theta y^r+\bar{G^\theta}
\end{equation}
\begin{equation}
G^\phi=\frac{1}{r}y^\phi y^r+\bar{G^\phi}.
\end{equation}

Here the prime indicates the derivative with respect to $r$, and
$\bar{G}^i$ are calculated from $\bar{F}^2$. Following Akbar-Zadeh
\cite{Akbar1988}, one can calculate Ricci tensor in Finsler
geometry from $Ric$ as
\begin{equation}
Ric_{\mu\nu}=\frac{\partial^2(\frac{1}{2}F^2Ric)}{\partial
y^\mu\partial y^\nu}.
\end{equation}

Also one can define the scalar curvature in Finsler   as
$S=g^{\mu\nu}Ric_{\mu\nu}$ and as a consequence, the modified
Einstein tensor in Finsler spacetime can be obtained  as
\begin{equation}
G_{\mu\nu}\equiv Ric_{\mu\nu}-\frac{1}{2}g_{\mu\nu}S
\end{equation}

Considering  $\bar{F}$ as dimensional Finsler spacetime with
constant flag curvature $\lambda$, one can find Einstein tensors
in Finsler geometry  as
\begin{equation}
G^t_t=\frac{A'}{rA^2}-\frac{1}{r^2A}+\frac{\lambda}{r^2},
\end{equation}

\begin{equation}
G^r_r=-\frac{B'}{rAB}-\frac{1}{r^2A}+\frac{\lambda}{r^2},
\end{equation}

\begin{equation}
G^\theta_\theta=G^\phi_\phi=-\frac{B''}{2AB}-\frac{B'}{2rAB}+\frac{A'}{2rA^2}+\frac{B'}{4AB}\left(\frac{A'}{A}+\frac{B'}{B}\right).
\end{equation}

As the matter distribution for constructing wormhole is still a
challenging issue to the physicists, we assume therefore, the
general anisotropic energy-momentum tensor~\cite{Rahaman2010} in
the form
\begin{equation}
 T_\nu^\mu=(\rho + p_r)u^{\mu}u_{\nu} + p_r g^{\mu}_{\nu}+
            (p_r -p_t )\eta^{\mu}\eta_{\nu},
\end{equation}
where $u^{\mu}u_{\mu} = - \eta^{\mu}\eta_{\mu} = 1$, $p_t$ and
$p_r$ are the transverse and radial pressures, respectively.

Using the above Finsler structure (1) and energy stress tensor
(22), one can write the gravitational field equations in the
Finsler geometry $(G^\mu_\nu=8\pi_FGT^\mu_\nu)$ as
\begin{equation}
8\pi_F G \rho=\frac{A'}{rA^2}-\frac{1}{r^2A}+\frac{\lambda}{r^2},
\end{equation}

\begin{equation}
-8\pi_F Gp_r=-\frac{B'}{rAB}-\frac{1}{r^2A}+\frac{\lambda}{r^2},
\end{equation}

\begin{equation}
-8\pi_F
Gp_t=-\frac{B''}{2AB}-\frac{B'}{2rAB}+\frac{A'}{2rA^2}+\frac{B'}{4AB}\left(\frac{A'}{A}+\frac{B'}{B}\right).
\end{equation}

Note that the $Ric$ from which the field equations are derived is
not dependent on connections, i.e. it is insensitive to the
connections. Secondly the  field equations  can be derived from a
Lagrangian approach. One can notice also that $\lambda$ which is
the beta part of the Finsler space fundamental function appears in
the field equations gives the Finslerian contribution. It is
important   to take into account the Cartan's connection approach
which  is the most convectional for studying gravitation field
equations in the framework of general relativity and gravitation.
The meaning is given in the fact of metrical connection
($g_{kl:m}$) which preserves the angle of two vectors moving along
the geodesics and the norm~\cite{Carroll}. It is a basic point in
the derivation of gravitation Einstein's equations. The
application of Cartan $d$-connection presents a difficulty to the
solutions of gravitational field. We avoid this approach in this
study, however, such an approach is possible.

To search for the wormhole solution we follow the convention given
by Morris and Thorne~\cite{MT1988} and hence write the above
equations in terms of the redshift function ($f(r)$) and shape
function ($b(r)$) by substituting $B(r) = e^{2f(r)}$ and $A(r) =
\frac{1}{1-\frac{b(r)}{r}}$. Thus the field equations (23)-(25)
take the following relationships
\begin{equation}
b' +\lambda -1 = 8\pi_F r^2G\rho,
\end{equation}

\begin{equation}
\left(1-\frac{b}{r}\right) \left(\frac{2f'}{r}+\frac{1}{r^2}\right)-\frac{\lambda}{r^2} = 8\pi_F G p_r,
\end{equation}

\begin{equation}
 \left\{1-\frac{b}{r}\right\}\left\{f''+\frac{f'}{r}+{f'}^2\right\}-\left \{\frac{b'}{r}-\frac{b}{r^2}\right
\}\left\{\frac{f'}{2}+\frac{1}{2r}\right\}= 8\pi_F  Gp_t.
\end{equation}

\section{Some models for wormholes}
 Einstein's general theory of relativity relates the matter distribution
with the geometry of the spacetime produced by the matter contain under
consideration. Thus if we know the geometry of the spacetime, then we can find the
corresponding matter distribution and vice versa. Also it has an
interesting feature that if one knows partly the geometry of the
spacetime and some components of energy stress tensor, then one can determine the total
structure of the spacetime as well as matter distribution through
field equations. Therefore, in the following text we are discussing
several models of the wormholes under different conditions.

 \subsection{Specific shape function and redshift function}
 In this subsection we assume some definite form of wormhole structures and try to find the
matter distributions that produce it.\\

\textbf{Case~1:} For particular shape function,  $ b(r)=r_0(\frac{r}{r_0})^n $, where, $r_0$ is the throat radius
and $n$ is an arbitrary constant, however, for satisfying flaring out, one has to take $n$ as less than unity~\cite{Lobo2006}.
Now, we shall consider two cases with different redshift functions: (i) $ f(r)= constant $, and (ii) $ f(r)=\frac{r_0}{r}$.
These two choices are justified as the redshift function $ f(r)$ must be finite for all values of $r$ to
avoid an event horizon.\\

{\it Subcase~(1a):} $f(r)= constant$\\
Using above field equations (26) - (28), we get the following stress-energy components:
\begin{equation}
\rho = \frac{n(\frac{r}{r_0})^{(n-1)}+(\lambda-1)}{8\pi_Fr^2G},
\end{equation}

\begin{equation}
p_r = \frac{-(\frac{r}{r_0})^{(n-1)}-(\lambda-1)}{8\pi_Fr^2G},
\end{equation}

\begin{equation}
p_t = -\frac{(n-1)(\frac{r_0}{r})^{(n-1)}}{16\pi_Fr^2G},
\end{equation}

\begin{equation}
\rho+p_r = \frac{(n-1)(\frac{r_0}{r})^{(n-1)}}{8\pi_Fr^2G}.
\end{equation}

{\it Subcase~(1b):} $f(r)=\frac{r_0}{r}$\\
Similarly, here we find the following stress-energy components:
\begin{equation}
\rho = \frac{n(\frac{r}{r_0})^{(n-1)}+(\lambda-1) }{8\pi_Fr^2G},
\end{equation}

\begin{equation}
p_r =\frac{2(\frac{r}{r_0})^{(n-2)}-(\frac{r}{r_0})^{(n-1)}-\frac{2r_0}{r}+1-\lambda}{8\pi_Fr^2G},
\end{equation}

\begin{equation}
p_t= \frac{-\frac{n-1}{2}(\frac{r}{r_0})^{(n-1)}+\frac{r_0}{r}
+(\frac{r_0}{r})^2-(\frac{r}{r_0})^{(n-3)}+\frac{n-3}{2}(\frac{r}{r_0})^{(n-2)}}{8\pi_Fr^2G},
\end{equation}

\begin{equation}
\rho+p_r =\frac{2(\frac{r}{r_0})^{(n-2)}+(n-1)(\frac{r}{r_0})^{(n-1)}-\frac{2r_0}{r}}{8\pi_Fr^2G}.
\end{equation}

\begin{figure*}
\begin{tabular}{rl}
\includegraphics[width=8.5cm]{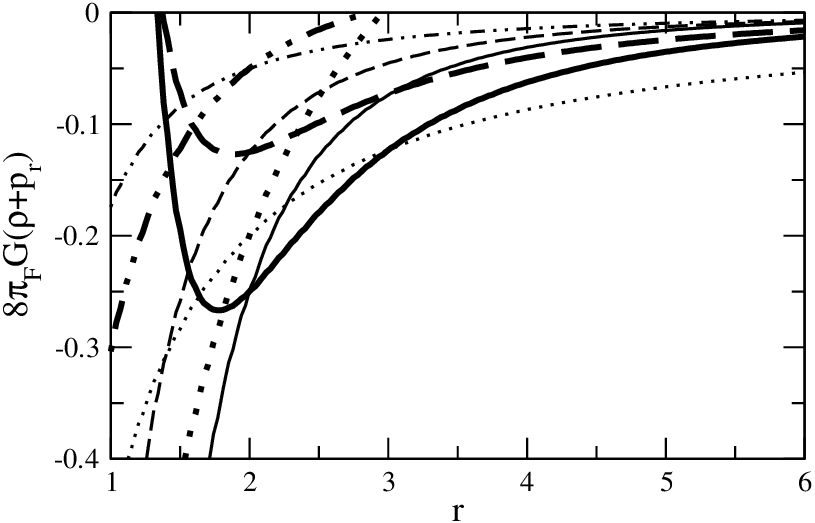}
\end{tabular}
\caption{Plot showing $\rho+p_r <0$ for case 1. Here $n=0.0$, $0.2$, $0.5$ and $0.8$ represented
by solid, dotted, dashed and chain curves, respectively. Thick curves represent $f(r)=r_0/r $
and thin curves represent f(r)=constant. We have assumed $r_0=2$ for the throat of the wormhole}
\end{figure*}

\textbf{Case~2:} We choose the shape function, $ b(r)=r_0+\rho_0r_0^3 \ln(\frac{r_0}{r})$, where
$r_0$ is the throat radius and $\rho_0$ is an arbitrary constant.  However, for satisfying the flare out condition,
one has to take $\rho_0$  as less than unity. We shall consider as above  two cases with
different redshift functions:\\

{\it Subcase~(2a):} $ f(r)= constant$\\
We find unknown parameters
\begin{equation}
\rho=\frac{(\lambda-1)r-\rho_0r_0^3}{8\pi_Fr^3G},
\end{equation}

\begin{equation}
p_r=\frac{(1-\lambda)r-[r_0+\rho_0r_0^3ln(\frac{r_0}{r})]}{8\pi_Fr^3G},
\end{equation}

\begin{equation}
p_t=\frac{r_0+\rho_0r_0^3[1+ln(\frac{r_0}{r})]}{16\pi_Fr^3G},
\end{equation}

\begin{equation}
\rho +p_r=-\frac{r_0+\rho_0r_0^3[1+ln(\frac{r_0}{r})]}{8\pi_Fr^3G}.
\end{equation}
\\

{\it Subcase~(2b):} $f(r)=\frac{r_0}{r}$\\
We obtain the unknown parameters as follows:
\begin{equation}
\rho=\frac{(\lambda-1)r-\rho_0r_0^3}{8\pi_Fr^3G},
\end{equation}

\begin{equation}
p_r=\frac{r(1-\lambda)-3r_0+\frac{2r_0^2}{r}+\rho_0r_0^3ln(\frac{r_0}{r})(\frac{2r_0}{r}-1)}{8\pi_Fr^3G},
\end{equation}

\begin{equation}
p_t=\frac{\frac{3r_0}{2}-\frac{r_0^3}{r^2}-\frac{r_0^2}{2r}+\rho_0
r_0^3\left[\frac{1}{2}-\frac{r_0}{2r}+ln(\frac{r_0}{r})\left(\frac{1}{2}-\frac{3r_0}{2r}-\frac{r_0^2}{r^2}\right)\right]}{8\pi_Fr^3G},
\end{equation}

\begin{equation}
\rho+p_r =\frac{-3r_0+\frac{2r_0^2}{r}+\rho_0r_0^3[ln(\frac{r_0}{r})(\frac{2r_0}{r}-1)-1]}{8\pi_Fr^3G}.
\end{equation}

\begin{figure*}
\begin{tabular}{rl}
\includegraphics[width=9.0cm]{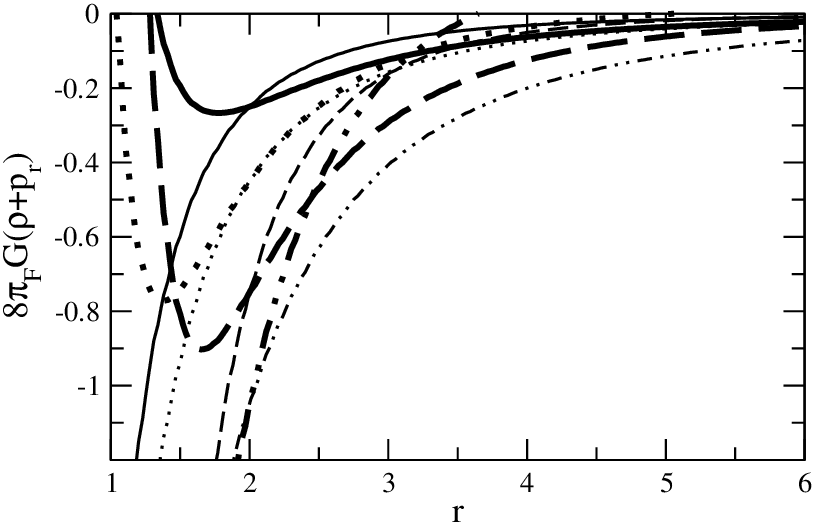}
\end{tabular}
\caption{
Plot showing $\rho+p_r <0$  for case 2. Here $\rho_0 = 0.0$, $0.2$, $0.5$ and $0.8$ represented by
solid, dotted, dashed and chain curves, respectively. Thick curves represent $f(r)=r_0/r $
and thin curves represent f(r)=constant. We have assumed $r_0=2$ for the throat of the wormhole}
\end{figure*}

\textbf{Case~3:} For the  shape function, $ b(r)=r_0+\gamma r_0(1-\frac{r_0}{r})$, where $r_0$
is the throat radius and $\gamma$ is an arbitrary constant, however, for satisfying  the flare out condition,
one has to take $\gamma$ as less than unity. We shall consider here also two cases with
different redshift functions: \\

{\it Subcase~(3a):} $f(r) = constant$\\
We obtain the unknown parameters
\begin{equation}
\rho= \frac{(\lambda-1)r+\gamma \frac{r_0^2}{r}}{8\pi_Fr^3G},
\end{equation}

\begin{equation}
p_r=\frac{(1-\lambda)r-r_0(1+\gamma )+\gamma \frac{r_0^2}{r}}{8\pi_Fr^3G},
\end{equation}

\begin{equation}
p_t=\frac{-2\gamma\frac{r_0^2}{r}+r_0(1+\gamma)}{16\pi_Fr^3G},
\end{equation}

\begin{equation}
\rho+p_r=\frac{2\gamma\frac{r_0^2}{r}-r_0(1+\gamma)}{8\pi_Fr^3G}.
\end{equation}

{\it Subcase~(3b):} $f(r)=\frac{r_0}{r}$\\
We obtain the unknown parameters
\begin{equation}
\rho= \frac{(\lambda-1)r+\gamma \frac{r_0^2}{r}}{8\pi_Fr^3G},
\end{equation}

\begin{equation}
p_r=\frac{r(1-\lambda)-3r_0+\frac{2r_0^2}{r}+\gamma
r_0(1-\frac{r_0}{r})(\frac{2r_0}{r}-1)}{8\pi_Fr^3G},
\end{equation}

\begin{equation}
p_t=\frac{\frac{3r_0}{2}-\frac{r_0^3}{r^2}-\frac{r_0^2}{2r}+\frac{\gamma
r_0^2}{r}\left[\frac{r_0}{2r}-\frac{1}{2}\right]+\gamma
r_0\left[1-\frac{r_0}{r}\right]\left[\frac{1}{2}-\frac{3r_0}{2r}-\frac{r_0^2}{r^2}\right]
}{8\pi_Fr^3G},
\end{equation}

\begin{equation}
\rho+p_r=\frac{\gamma \frac{r_0^2}{r}-3r_0+\frac{2r_0^2}{r}+\gamma
r_0(1-\frac{r_0}{r})(\frac{2r_0}{r}-1)}{8\pi_Fr^3G}.
\end{equation}


\begin{figure*}
\begin{tabular}{rl}
\includegraphics[width=9.0cm]{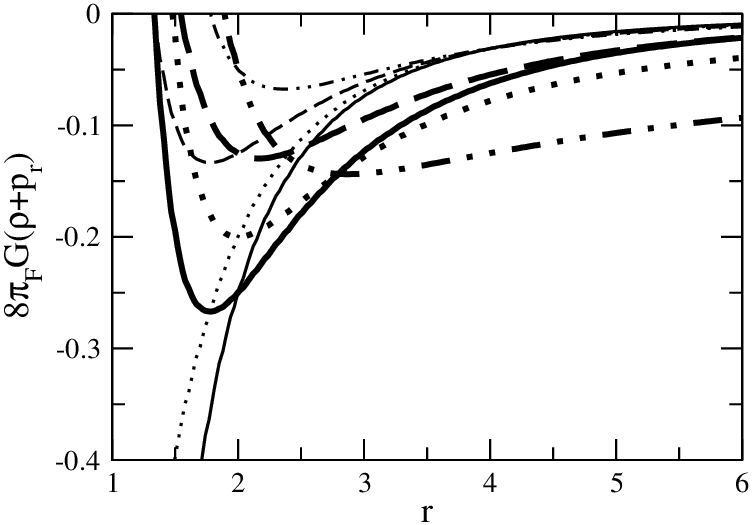}
\end{tabular}
\caption{
Plot showing $\rho+p_r <0$ for case 3. Here $\gamma =0.0$, $0.2$, $0.5$ and $0.8$ represented
by solid, dotted, dashed and chain curves, respectively. Thick curves represent $f(r)=r_0/r $
and thin curves represent f(r)=constant. We have assumed $r_0=2$ for the throat of the wormhole}
\end{figure*}


\subsection{Specific energy density and redshift function}

\textbf{Case~4:} For the specific energy density, $\rho = \rho _0(\frac{r_0}{r})^\alpha$, where
$r_0$, $\rho_0$ and $\alpha$ are arbitrary constants, we shall consider two cases with
different redshift functions:\\

{\it Subcase~(4a):} $f(r) = constant$\\
Using the above choices of energy density and redshift function, we obtain the shape function
$b(r)$ from field Eq. (26) as
\begin{equation}
b=c_1-\left[(\lambda-1)+\frac{8\pi_F r^2G\rho_0}{\alpha-3}\left(\frac{r_0}{r}\right)^\alpha\right]r,
\end{equation}
where $c_1 $ is an integration constant.

The radial and transverse pressures are obtained as
\begin{equation}
8\pi_F r^2Gp_r=\frac{8\pi_F r^2G\rho_0}{\alpha-3}\left(\frac{r_0}{r}\right)^\alpha-\frac{c_1}{r},
\end{equation}

\begin{equation}
16\pi_F r^2Gp_t=\left[\frac{c_1}{r}-\left( \frac{\alpha-2}{\alpha-3} \right)8\pi_F r^2G\rho_0\left(\frac{r_0}{r}\right)^\alpha\right],
 \end{equation}

\begin{equation}
8\pi_F r^2G(\rho+p_r)=\left(\frac{\alpha-2}{\alpha-3}\right)8\pi_F r^2G\rho_0\left(\frac{r_0}{r}\right)^\alpha-\frac{c_1}{r}.
\end{equation}

{\it Subcase~(4b):} $ f(r)=\frac{r_0}{r}$\\
We obtain the unknown parameters
\begin{equation}
b=c_1-\left[(\lambda -1)+\frac{8\pi_F r^2G\rho_0}{\alpha-3}\left(\frac{r_0}{r}\right)^\alpha\right]r,
\end{equation}

\begin{equation}
8\pi_F r^2Gp_r=\left[\frac{8\pi_F r^2G\rho_0}{\alpha-3}\left(\frac{r_0}{r}\right)^\alpha-\frac{c_1}{r}\right]\left[1-\frac{2r_0}{r}\right]-2\lambda
\frac{r_0}{r},
\end{equation}

\begin{equation}
8\pi_F r^2Gp_t=\left[\frac{8\pi_F r^2G\rho_0}{\alpha-3}\left(\frac{r_0}{r}\right)^\alpha
-\frac{c_1}{r}\right]\left[\left(\frac{r_0}{r}\right)^2+\frac{r_0}{r}\right],
\end{equation}

\begin{equation}
+\lambda \left[\left(\frac{r_0}{r}\right)^2+\frac{r_0}{r}\right]
-\left[\frac{\alpha-2}{\alpha-3}8 \pi_F r^2 G \rho_0
\left(\frac{r_0}{r}\right)^\alpha - \frac{c_1}{r}
\right]\left[\frac{1}{2}-\frac{r_0}{2r}\right],
\end{equation}

\begin{equation}
8\pi_F r^2G(\rho+p_r)=\left[\frac{8\pi_F r^2G \rho_0}{\alpha-3}\left(\frac{r_0}{r}\right)^\alpha
-\frac{c_1}{r}\right] \left[1-\frac{2r_0}{r}\right]-2\lambda \frac{r_0}{r} + 8\pi
r^2G\rho_0\left(\frac{r_0}{r}\right)^\alpha.
\end{equation}

Note that $ b(r)$ has the same form as the case $f = constant$. Therefore, we  have same  plots for
$b(r)$, $b(r) - r$ and $b'(r)$ when $f = constant $.


\begin{figure*}
\begin{tabular}{rl}
\includegraphics[width=9.0cm]{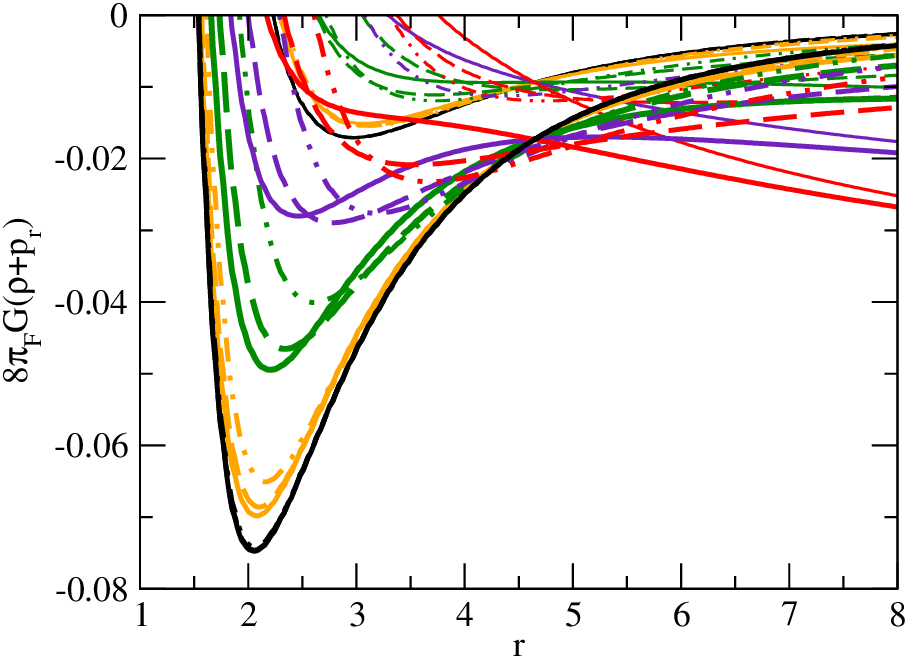}
\end{tabular}
\caption{Plot showing $\rho+p_r <0$ for {\it Case 4}. Here, $ \rho_0 = 0.001$, $0.01$, $0.05$, $0.10$ and $0.15$
are represented by black, orange, green, indigo and red colours, respectively. $\alpha $ is taken to be $0.0$, $1.0$
and $2.0$ are represented by solid, dashed and chain curves, respectively. Thick curves represent $f(r)=r_0/r $
and thin curves represent $f(r)=constant$. We have assumed $r_0=2$ and $c_1=1$}
\end{figure*}



\begin{figure*}
\begin{tabular}{rl}
\includegraphics[width=8.0cm]{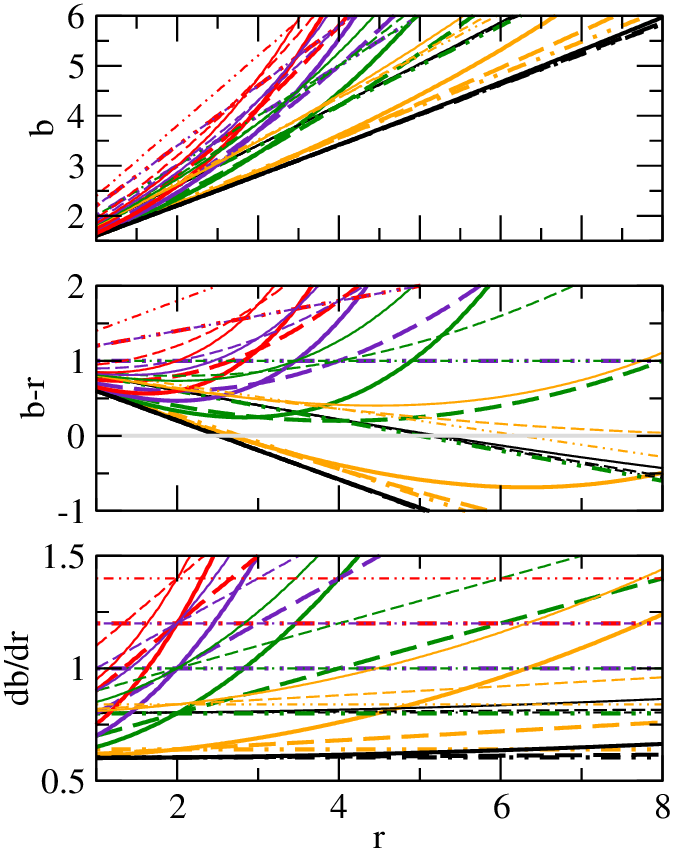}
\end{tabular}
\caption{Plots (upper, middle and lower panels respectively) showing the behavior of the shape function, radii of the
throat where $b-r$ cuts $r$ axis and nature of derivative of the  shape function for {\it Case 4}. Here,
$\rho_0 = 0.001$, $0.01$, $0.05$, $0.10$ and $0.15$ are represented by black, orange, green, indigo and red colours,
respectively. $\alpha $ is taken to be $0.0$,  $0.5$, $1.0$, $1.5$ and $2.0$, represented by solid, dotted,
dashed, dot-dashed and chain curves, respectively. Thick curves represent $f(r)=r_0/r $ and thin curves represent
$f(r)=constant$. We have assumed $r_0=2$ and $c_1=1$}
\end{figure*}


\textbf{Case~5:} For the dark energy equation of state, $p_r=\omega \rho;~\omega<-1$, we shall consider as above
two cases with different redshift functions:\\

{\it Subcase~(5a):} $f(r)=$constant\\
Using the above choices of energy density and redshift function, we obtain the following parameters
\begin{equation}
b=(1-\lambda)r+r_0\left(\frac{r_0}{r}\right)^{(\frac{1}{\omega})},
\end{equation}

\begin{equation}
 8\pi_F r^2G \rho=-\frac{1}{\omega}\left(\frac{r_0}{r}\right)^{(\frac{1}{\omega}+1)},
\end{equation}

\begin{equation}
 8\pi_F r^2Gp_r=-\left(\frac{r_0}{r}\right)^{(\frac{1}{\omega}+1)},
\end{equation}

\begin{equation}
8\pi_Fr^2Gp_t=\frac{1}{2}\left(\frac{1}{\omega}+1\right)\left(\frac{r_0}{r}\right)^{(\frac{1}{\omega}+1)},
\end{equation}
where $r_0 ^{(\frac{1}{\omega}+1)}$ is an integration constant.\\


\begin{figure*}
\begin{tabular}{rl}
\includegraphics[width=9.0cm]{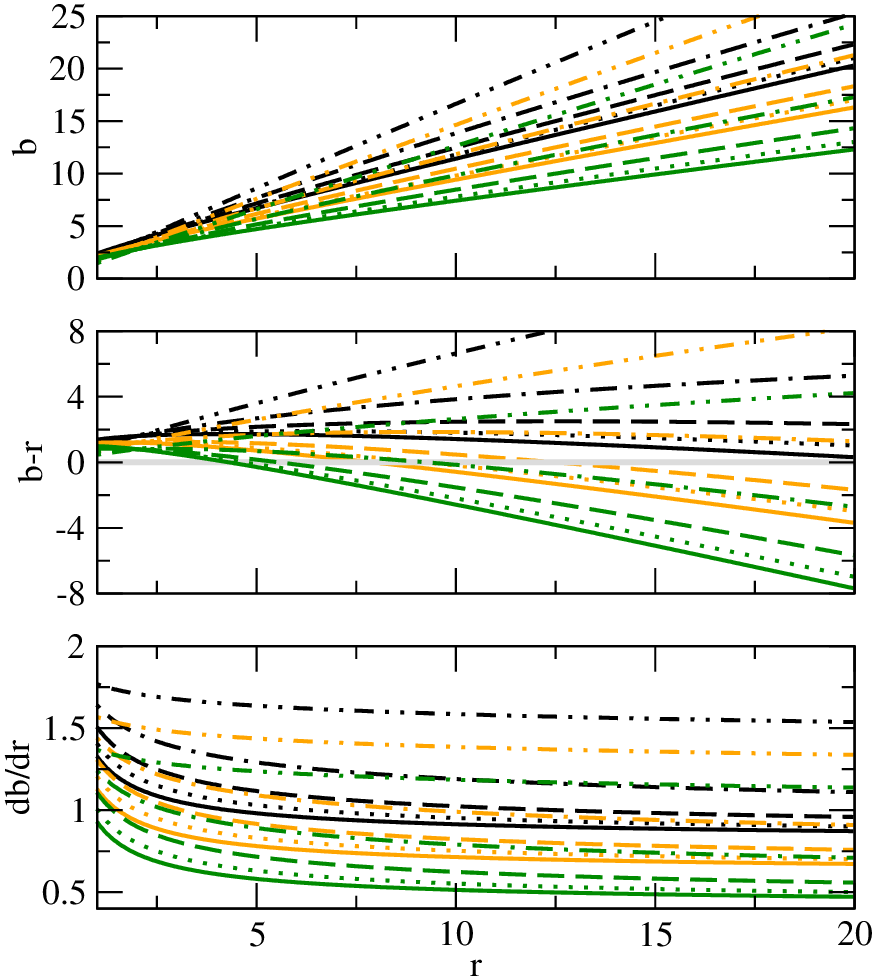}
\end{tabular}
\caption{Plots showing the behavior of the shape function, radii of the throat where $b-r$ cuts $r$ axis
and nature of derivative of the shape function for {\it Case 5a}. Here, $\lambda= 0.2$, $0.4$ and $0.6$
represented by black, orange and green colors. Solid, dotted, dashed, dot-dashed and chain curves represent
$\omega=-3.0$, $-2.5$, $-2.0$, $-1.5$, $-1.1$, respectively. We have assumed $r_0=2$ for the throat of the wormhole}
\end{figure*}


{\it Subcase~(5b):}  $f(r)=\frac{r_0}{r}$\\
We obtain the unknown parameters as follows:
\begin{equation}
b=(1-\lambda)r+\left[\frac{4\lambda r_0^2}{\omega r} \right]
\left[ \left(-\frac{2r_0}{\omega r} \right)^{\left(
\frac{1}{\omega}-1\right)}\exp\left(-\frac{2r_0}{\omega r
}\right)\left\{\Gamma \left(1-\frac{1}{\omega}\right)-\Gamma
\left(1-\frac{1}{\omega},-\frac{2r_0}{\omega
r}\right)\right\}-\frac{\omega r}{2r_0} \right]+c_3,
\end{equation}

\begin{equation}
8\pi_Fr^2G\omega
\rho=(1-\lambda)+\left(\frac{2br_0}{r^2}-\frac{2r_0}{r}\right)-\frac{b}{r},
\end{equation}

\begin{equation}
8\pi_Fr^2Gp_r=(1-\lambda)+\left(\frac{2br_0}{r^2}-\frac{2r_0}{r}\right)-\frac{b}{r},
\end{equation}

\begin{eqnarray}
8\pi_Fr^2Gp_t=\left[1-\frac{b}{r}\right]\left[\frac{r^2_0}{r^2}+\frac{r_0}{r}\right]-\left[
(1-\lambda)\left(1+\frac{1}{\omega}\right)+\frac{1}{\omega}\left\{\frac{2br_0}{r^2}-
\frac{2r_0}{r}\right\}-\left(1+\frac{1}{\omega}\right)\frac{b}{r}
 \right] \nonumber \\ \times \left[\frac{1}{2}-\frac{r_0}{2r}\right],
\end{eqnarray}
where $c_3 $ is an integration constant.


\begin{figure*}
\begin{tabular}{rl}
\includegraphics[width=9.0cm]{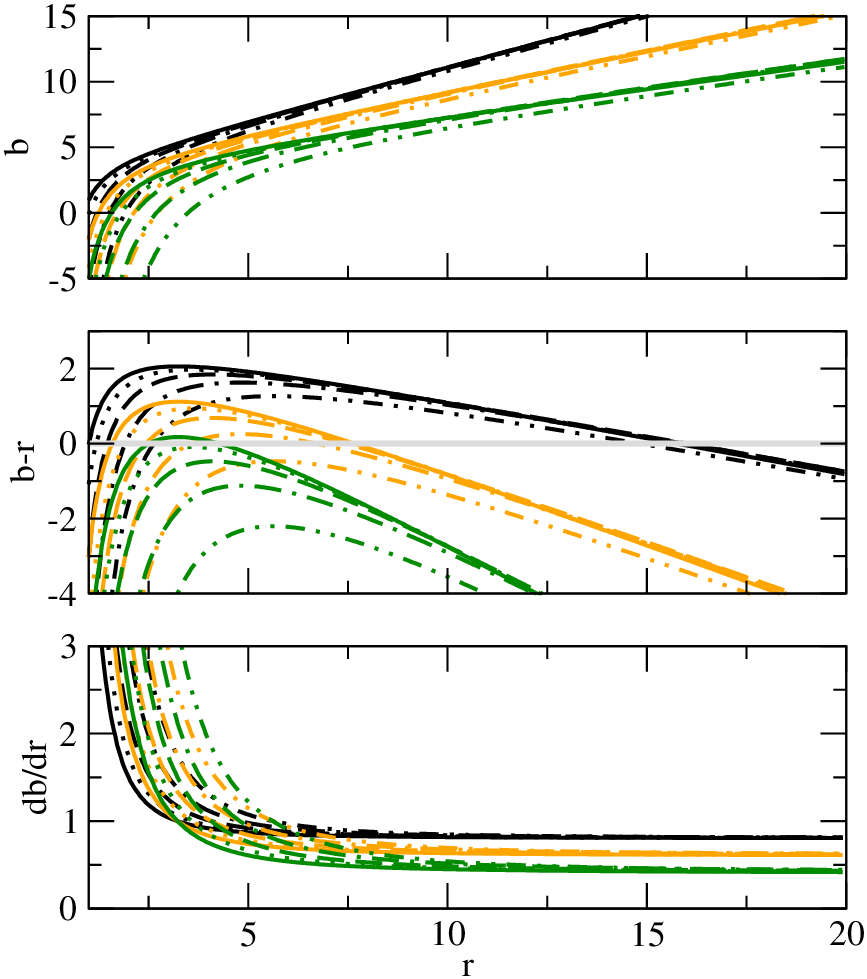}
\end{tabular}
\caption{Plots showing the behavior of the shape function, radii of the throat where $b-r$ cuts $r$ axis
and nature of derivative of the  shape function for {\it Case 5b}. Here $\lambda= 0.2$, $0.4$ and $0.6$, represented
by black, orange and green colors. Solid, dotted, dashed, dot-dashed and chain curves represent
$\omega=-3.0$, $-2.5$, $-2.0$, $-1.5$, $-1.1$, respectively. We have assumed $r_0=2$ and $c_3=3$}
\end{figure*}


\section{Discussion and Conclusion }
Recent literature survey exhibits that the Finsler geometry has
accumulated much attention due to its potentiality to explain
various issues, specially as cosmic acceleration, which can be
explained without invoking dark matter~\cite{Chang2008} or dark
energy~\cite{Chang2009}. In the context of GR, the violation of
NEC (often called {\it exotic matter}) is a basic ingredient of
static traversable wormholes, although the violation of energy
conditions is quite acceptable in certain quantum fields, among
which the Casimir effect and Hawking evaporation are mentionable.
The present work may be looked upon as a possible solutions to
construct theoretically traversable wormholes in the context of
Finsler geometry. In this context, we derived the Einstein
gravitational field equations in the Riemannian spacetime with
matter distribution as anisotropic in nature. We find out our
solution in conventional way like the Morris-Thorne wormhole
solution. We focus our attention mainly for discussing the
violation of null energy condition (NEC) and the constraint on
wormhole geometry, respectively.

In the present work we obtain exact solutions by imposing
restricted choices of the redshift function, the shape function
and/or specifying an equations of state. Some of the important
features of the present investigation can be formulated as
follows:

(i) In first three cases we consider the various choices for the
form function, namely, $ b(r)=r_0(\frac{r}{r_0})^n $, $
b(r)=r_0+\rho_0 r_0^3 \ln(\frac{r_0}{r})$ and $b(r)=r_0+\gamma
r_0(1-\frac{r_0}{r})$ and we have analyzed the solution by
considering that the redshift function can either be constant, or
have the functional relation of radial co-ordinate.

(ii) In the next two cases we have found out the shape functions
for the specific form of energy density, namely, $\rho = \rho _0(\frac{r_0}{r})^\alpha$
and dark energy equation of state, $p_r=\omega \rho;~\omega<-1$.

After knowing all the metric potentials $f(r)$, $b(r)$ and
stress-energy components $\rho$, $p_r$ and $p_t$, we examine
whether the results indeed give wormhole structures. It is
essentially required that to get a wormhole, the following
properties must be satisfied:\\

1) The \emph{redshift function}, $f(r)$, should remain finite everywhere to
prevent an event horizon.

2) The \emph{shape function}, $b(r)$, should obey the following flare-out conditions at the
 throat $r = r_0$ : $b(r_0) = r_0$ and $b^\prime(r_0) < 1$, $r_0$ being the throat radius.

3) Another condition that needed to satisfy is $b(r)/r < 1$ for $r >r_0$.

4) The NEC must be violated for traversable wormhole, i.e. $p_r
+\rho < 0 $.

The first three conditions for the geometry of the spacetime and
last one for the matter distribution that produces this spacetime.

We have, however, verified whether our models satisfy all the
criteria to represent wormhole structure as follows:

In models 1-3, we have assumed that the spacetime produces
wormholes and try to search for the matter distributions which
produce these features. We have found out the components of the
energy momentum tensors and Figs. 1-4 indicate the matter
distributions violate the NEC. Note that violation of NEC is one
of the important criteria to hold a wormhole open. Thus the first
three models are physically valid. On the other hand, in models 4
and 5, we have found out the shape functions for the specific form
of energy density and dark energy equation of state. For model 4,
one can note that the shape function $b(r)$ assumes the same form
for constant or specific redshift function. In Fig. 5, we have
vividly depicted different characteristics of the shape function.
We observe that existence of the throat depends on the choices of
the parameters. The radius of the throat exists where $b(r) - r$
cuts $r$ axis. Also this figure indicates that flaring out
condition is satisfied at the throat i.e. $b'(r_0)<1$. Thus in
the model 4, the above four conditions are satisfied for
development of wormholes structure.

We have also analyzed, in the model 5, different characteristics
of the shape functions for different redshift functions in Figs. 6
and 7. These figures satisfy all the geometric criteria of the
wormhole structure. In this case we have assumed dark energy
equation of state, $p_r=\omega \rho;~\omega<-1$, and hence the
NEC, $p_r +\rho < 0$ is automatically satisfied. Thus, it is an
overall observation that present models successfully describe the
wormhole features under the  background of the Finslerian
spacetime.

\subsection*{Acknowledgments}
FR, SR and AAU are thankful to IUCAA for providing Associateship
under which a part the work was carried out. AB is also grateful
to IUCAA for providing research facilities and hospitality. FR is
thankful to DST, Govt. of India for providing financial support
under PURSE programme. Finally we are grateful to the referee for
several valuable comments and suggestions which have improved the
manuscript substantially.

\end{document}